\documentclass[12pt]{article}
\usepackage{amsfonts}
\usepackage{amsmath, amsthm}
\usepackage{epsfig}
\usepackage{rotating}
\usepackage{amsmath,empheq}
\usepackage{authblk}
\usepackage{mathtools}

\usepackage[toc,page]{appendix}

\usepackage{algorithm,algorithmic}
\usepackage{bm}

\usepackage{authblk}

\usepackage{amsmath,verbatim,color,amssymb,epsfig}
\usepackage{bm}
\usepackage{amsfonts}
\usepackage{subfig}
\usepackage{epsfig}
\usepackage{multirow}
\usepackage{graphicx}
\usepackage{array}
\usepackage[table]{xcolor}
\definecolor{Gray}{gray}{0.80}
\usepackage{lscape}
\usepackage{cite}
\usepackage{natbib}
\usepackage[normalem]{ulem}
\usepackage[colorlinks=true,urlcolor=blue,citecolor=purple,linkcolor=blue,bookmarks=true]{hyperref}
\DeclareMathOperator*{\argmax}{argmax} 

\usepackage{caption}
\begin{document}
\def\eqx"#1"{{\label{#1}}}
\def\eqn"#1"{{\ref{#1}}}
\newtheorem{remark}{Remark}
\makeatletter 
\@addtoreset{equation}{section}
\makeatother  

\def\yincomment#1{\vskip 2mm\boxit{\vskip 2mm{\color{red}\bf#1} {\color{blue}\bf --Yin\vskip 2mm}}\vskip 2mm}
\def\squarebox#1{\hbox to #1{\hfill\vbox to #1{\vfill}}}
\def\boxit#1{\vbox{\hrule\hbox{\vrule\kern6pt
          \vbox{\kern6pt#1\kern6pt}\kern6pt\vrule}\hrule}}

\def\theequation{\thesection.\arabic{equation}}
\newcommand{\ds}{\displaystyle}

\newcommand{\bJ}{\mbox{\bf J}}
\newcommand{\bF}{\mbox{\bf F}}
\newcommand{\bM}{\mbox{\bf M}}
\newcommand{\bR}{\mbox{\bf R}}
\newcommand{\bZ}{\mboxZ}
\newcommand{\bX}{\mbox{\bf X}}
\newcommand{\bx}{\mbox{\bf x}}
\newcommand{\bQ}{\mbox{\bf Q}}
\newcommand{\bH}{\mbox{\bf H}}
\newcommand{\bh}{\mbox{\bf h}}
\newcommand{\bz}{\mboxZ}
\newcommand{\ba}{\mbox{\bf a}}
\newcommand{\be}{\mbox{\bf e}}
\newcommand{\bG}{\mboxG}
\newcommand{\bB}{\mbox{\bf B}}
\newcommand{\bb}{\mbox{\bf b}}
\newcommand{\bA}{\mbox{\bf A}}
\newcommand{\bC}{\mbox{\bf C}}
\newcommand{\bI}{\mbox{\bf I}}
\newcommand{\bD}{\mbox{\bf D}}
\newcommand{\bU}{\mbox{\bf U}}
\newcommand{\bc}{\mbox{\bf c}}
\newcommand{\bd}{\mbox{\bf d}}
\newcommand{\bs}{\mbox{\bf s}}
\newcommand{\bS}{\mbox{\bf S}}
\newcommand{\bV}{\mbox{\bf V}}
\newcommand{\bv}{\mbox{\bf v}}
\newcommand{\bW}{\mbox{\bf W}}
\newcommand{\bw}{\mbox{\bf w}}
\newcommand{\bg}{\mboxG}
\newcommand{\bu}{\mbox{\bf u}}
\def\bb{{\bf b}}

\newcommand{\bcU}{\boldsymbol{\cal U}}
\newcommand{\bbeta}{\boldsymbol{\beta}}
\newcommand{\bdelta}{\boldsymbol{\delta}}
\newcommand{\bDelta}{\boldsymbol{\Delta}}
\newcommand{\boldeta}{\boldsymbol{\eta}}
\newcommand{\bxi}{\boldsymbol{\xi}}
\newcommand{\bGamma}{\boldsymbol{\Gamma}}
\newcommand{\bSigma}{\boldsymbol{\Sigma}}
\newcommand{\balpha}{\boldsymbol{\alpha}}
\newcommand{\bOmega}{\boldsymbol{ R}}
\newcommand{\btheta}{\boldsymbol{\theta}}
\newcommand{\bmu}{\boldsymbol{\mu}}
\newcommand{\bnu}{\boldsymbol{\nu}}
\newcommand{\bgamma}{\boldsymbol{\gamma}}

\newtheorem{thm}{Theorem}[section]
\newtheorem{lem}{Lemma}[section]
\newtheorem{rem}{Remark}[section]
\newtheorem{cor}{Corollary}[section]
\newcolumntype{L}[1]{>{\raggedright\let\newline\\\arraybackslash\hspace{0pt}}m{#1}}
\newcolumntype{C}[1]{>{\centering\let\newline\\\arraybackslash\hspace{0pt}}m{#1}}
\newcolumntype{R}[1]{>{\raggedleft\let\newline\\\arraybackslash\hspace{0pt}}m{#1}}

\newcommand{\tabincell}[2]{\begin{tabular}{@{}#1@{}}#2\end{tabular}}
\def\correspondingauthor{\footnote{hl3213@columbia.edu}}

\title{\bf Reinforcement Learning: \\Prediction, Control and Value Function Approximation}

\author[1]{Haoqian Li}
\author[2]{Thomas Lau}

\affil[1]{Columbia University}
\affil[2]{Point Zero One Technology}

\maketitle
\begin{abstract}
With the increasing power of computers and the rapid development of self-learning methodologies such as machine learning and artificial intelligence, the problem of constructing an automatic Financial Trading Systems (FTFs) becomes an increasingly attractive research topic. An intuitive way of developing such a trading algorithm is to use Reinforcement Learning (RL) algorithms, which does not require model-building. In this paper, we dive into the RL algorithms and illustrate the definitions of the reward function, actions and policy functions in details, as well as introducing algorithms that could be applied to FTFs. 

\end{abstract}
\section{Introduction}
In quantitative trading, a trader's objective is to optimize some measure of the performance of the executions, e.g., profit or risk-adjusted return, subjected to some certain constraints. There is a lot of work using predictive models of price changes for quantitative trading, especially for high-frequency trading and market microstructure data (\cite{Kearns2013}; \cite{Gerlein2016}; \cite{Schumaker2009} and \cite{Abis2017}). These models are trained based on specific machine learning objective functions (e.g., regression and classification loss functions), and thus there is no guarantee for the models to globally optimize their performances under the measure of the trader's objective.

The financial market is one of the most dynamic and fluctuating entities that exist, which makes it difficult to model its behavior accurately. However, the Reinforcement Learning algorithm, as another type of self-adaptive approach, can conquer these type of difficulties by directly learning from the outcomes of its actions. More specifically, the investment decision making in RL is a stochastic control problem, or a Markov Decision Process (MDP), where the trading strategies are learned from direct interactions with the market. Thus the need for building forecasting models for future prices or returns is eliminated.

Many research has been conducted within the field of RL in quantitative trading (\cite{Bertoluzzo2012}; \cite{Moody1998}; \cite{Moody2001}; \cite{Gold2003}; \cite{Nevmyvaka2006} and \cite{Eilers2014}). Generally, they show that the trading strategies based on RL perform better than those based on the supervised machine learning methodologies. Additionally, they acknowledged the ways of using Direct Recurrent RL (DDRL) approach instead of using the traditional TD-learning and Q-learning in the RL field. 

In this paper, we first present some basic aspects of RL in Section \ref{RL} and then go through the MDP and the reward function in Section \ref{MDP}. In Section \ref{Modelfree}, we introduce the model-free prediction and control, while in Section \ref{VFA}, we talk about the value function approximation. In Section \ref{AC}, we provide an algorithm that combines the knowledge of all previous sections.

\section{Basic Framework}\label{RL}
Reinforcement Learning (RL), rooted in the field of control theory, is a branch of machine learning explicitly designed for taking suitable action to maximize the cumulative reward. It is employed by an agent to take actions in an environment so as to find the best possible behavior or path it should take in a specific situation. Reinforcement learning differs from the traditional supervised learning as in supervised learning the training data has the answer key with it provided by an external supervisor, and the model is trained with the correct answer. Whereas in reinforcement learning, the reinforcement agent decides what to do to perform well (quantified by a defined reward function) in the given task. 

\section{MDP and The Value Function}\label{MDP}
RL lies in the interactions between the \emph{agent} and the \emph{environment}. At any time $t$, the agent receives input from the environment (observations) $O_t$, take some action (possibly random) $A_t$, and receive a reward (immediate/long-term) $R_t$ from the environment. 

\subsection{Basic Framework}
The \emph{history} is a sequence of observations, actions, and rewards $H_t=A_1,O_1, R_1,\dots, A_t, O_t, R_t,$ where the agent selects actions and the environment selects observations/rewards. The \emph{state} is the information determining what happens next. Formally, an information state is a function of the history $S_t=f(H_t),$ containing all the useful information from history, and the sequence of process is assumed to possess Markov property. 

An RL agent consists of two components, namely, \emph{policy} and \emph{value} function. A policy $\pi (a|s)=P[A=a|S=s]$ is a distribution over actions given the state $s$, which fully defines an agent's behavior mapping from state to action $\pi(S_t)=A_t.$ While the value function represents the goodness of each state based on the long-term expected cumulative rewards.

\subsection{Reward and Return}
A reward $R_t$ is a scalar feedback signal indicating how well an agent is doing at time $t.$ In the trading scenario, we can apply the Sharpe Ratio or the differential Sharpe Ratio proposed by \cite{Moody1998} as $R_t$, which is a better estimate for the risk-adjusted profit. The return $G_t$ is the total discounted reward starting from time-step $t$ 
\begin{eqnarray*}
G_t=R_{t+1}+\gamma R_{t+2}+\dots=\displaystyle {\sum^\infty_{k=0}} \gamma^k R_{t+k+1},
\end{eqnarray*} 
where $\gamma\in (0,1)$ is the discount factor. In some settings, a long-term reward is delayed, and it is better for the agent to sacrifice the immediate reward in exchange for the long-term reward. For instance, in the stock market, one can achieve long-term reward and gain more long-term risk-adjusted return by sacrificing the short-term stock return.

\subsection{MDP}
In the RL framework, it is usually assumed that the system satisfies the Markov property
$$P[S_{t+1}|S_t]=P[S_{t+1}|S_1,\dots,S_t],$$
which states the fact that the probability of transition from the current state $S_t$ to the next $S_{t+1}$ depends only on the current state $S_t$, instead of the whole history, i.e., the future is independent to the past given the present. MDP is an environment where all states are Markov and can be viewed as a tuple $\langle S,A,R,\gamma \rangle$ where $S$ is a finite set of states, $A$ is a finite set of actions, $R$ is a reward function for taking some action $a$ at state $s$ and time-step $t$ 
$$R_s^a=\mathbb E[R_{t+1}|S_t=s, A_t=a],$$ 
and $\gamma$ is the discount factor. Given the MDP and a policy $\pi$, the state sequence $S_1, S_2,\dots$ is a Markov process. 

\subsection{Value Function}
In RL, there are basically two types of value functions, namely, the expected return of a state, and the expected return of an action.
The state-value function $v_{\pi}(s)$ of an MDP measures the expected return starting from state $s$, given the policy $\pi$
\begin{eqnarray*}
v_\pi(s)=\mathbb E_\pi [G_t|S_t=s] 
=E_\pi[R_t+\gamma R_{t+1}+\gamma^2 R_{t+2}+...|S_t=s],
\end{eqnarray*}
while the action-value function $q_\pi(s,a)$ is the expected return starting from state $s$, taking action $a$, and following the policy $\pi$ 
$$q_\pi(s,a)=\mathbb E_\pi [G_t|S_t=s,A_t=a].$$
We can decompose the value functions into immediate reward plus discounted value of successor state
$$v_\pi(s)=\mathbb E_\pi [R_{t+1}+\gamma v_\pi(S_{t+1})|S_t=s],$$
$$q_\pi(s,a)=\mathbb E_\pi [R_{t+1}+\gamma q_\pi(S_{t+1},A_{t+1})|S_t=s, A_t=a],$$
which can be demonstrated as the Bellman Equations, which is a cornerstone of algorithms such as the TD-learning and Q-learning. 
A policy $\pi$ is said to outperform another $\pi^{'},$ if $v_\pi(s)\geqslant v_{\pi^{'}}(s),$ for $\forall s,$ and the agent's objective is to find an optimal policy that is better than or equal to all the other policies. 
The optimal policy identifies the values $v^*(s)$ and $q^*(s,a)$ such that 
\begin{center}$v^*(s)=\max\limits_{\pi} v_\pi(s)$ and $q^*(s,a)=\max\limits_{\pi} q_\pi(s,a), \forall (s, a).$\end{center} 
In fact, the optimal value functions are recursively determined by the Bellman optimality equations
\begin{center}$v^*(s)=\max\limits_a q^*(s,a),$\end{center} which states the fact that the value of a state under the optimal policy should be equal to the expected return for the optimal action from the state itself.

\section{Model-Free Prediction and Control}\label{Modelfree}
\subsection{Model-Free Prediction}
Model-free prediction learns an unknown MDP by estimating its value function and there are three approaches, namely, the Monte-Carlo learning (MC), Temporal-Difference Learning (TD(0)), and the TD($\lambda$). Since MC learns from complete episodes and requires all episodes to terminate, in this paper, we only focus on the TD(0) and TD($\lambda$) algorithm.

TD methods learn $v_\pi(s)$ directly from episodes of experience under policy $\pi$ and there is no requirement for the knowledge of MDP transitions and rewards. TD differs from MC in a way that TD learns from incomplete episodes by \emph{bootstrapping}. Denote $\mu_1,\mu_2,\dots$ as the mean of a sequence $x_1,x_2,\dots$ that can be computed incrementally. Then,
\begin{eqnarray*}
\mu_k&=&\frac{1}{k}\sum\limits_{j=1}^k x_j\\
&=&\frac{1}{k}(x_k+\sum\limits_{j=1}^{k-1} x_j)\\
&=&\frac{1}{k}(x_k+(k-1)\mu_{k-1})\\
&=&\mu_{k-1}+\frac{1}{k}(x_k-\mu_{k-1}).
\end{eqnarray*}
If we apply this to the value function $v_\pi(s)$ and treat the expected value as an empirical mean, then given a policy $\pi$, we can update the value function $V(S_t)$ toward the \emph{estimated return} $G_t=R_{t+1}+\gamma V(S_{t+1})$ by the following equation
\begin{center}
$V^{k+1}(S_t)= V^{k}(S_t)+\frac{1}{k+1}(R^{k+1}_{t+1}+\gamma V^{k+1}(S_{t+1})-V^{k}(S_t))$,  
\end{center}\begin{center}
or
$V(S_t)\leftarrow V(S_t)+\alpha(R_{t+1}+\gamma V(S_{t+1})-V(S_t)),$
\end{center} 
where $\alpha=\frac{1}{k+1}$. Note that the subtle difference $R_{t+1}+\gamma v_\pi(S_{t+1})$ is an unbiased estimate of $v_\pi(S_t),$ whereas $R_{t+1}+\gamma V(S_{t+1})$ is biased. This method is called TD(0) that essentially look at one-step further when adjusting value estimates.

Naturally, we can generalize the TD(0) algorithm to $n$ steps into the future, and define the n-step return as 
$$G_t^{(n)}=R_{t+1}+\gamma R_{t+2}+\dots+\gamma^{n-1}R_{t+n}+\gamma^nV(S_{t+n}),$$ 
which is the cumulative return of $n$ time-steps plus the value onward. 
We then average the n-step returns over different $n$ and treat them as our new $G_t,$ which combines all the n-step returns $G_t^{(n)}$. 
More specifically, we use weight $(1-\lambda)\lambda^{n-1}$ to define the $\lambda$-return $G_t^\lambda$ as 
$$G_t^\lambda=(1-\lambda)\sum\limits_{n=1}^\infty \lambda^{n-1}G_t^{(n)},$$ 
where $\lambda\in [0,1]$. When $\lambda=1$, the credit is deferred until the end of the episode (long-term); while when $\lambda=0$, the algorithm only looks at reward one step further (myopic).

Similar to the TD(0), we can use $G_t^\lambda$ to define the TD$(\lambda)$ algorithm, in which we update $V(S_t)$ as 
$$V(S_t)\leftarrow V(S_t)+\alpha(G_t^\lambda-V(S_t)),$$
which is called the forward-view of TD($\lambda$) as we update the value function towards the $\lambda-$return. 
However, this forward algorithm requires knowledge of the reward completely, and therefore can only be computed from complete episodes. 
To solve this problem, we use a backward TD($\lambda$) to update online from incomplete sequences by introducing the eligibility traces $E_t(s)$ of a state $s$, which is the degree to which it has been visited in the recent past including both frequency heuristic and recency heuristic. One version of the eligibility trace is defined as 
\begin{center}$E_t(s)=\sum\limits_{k=1}^t (\lambda \gamma)^{t-k}\textbf{1}(S_t=s).$\end{center}
which is used to update all the states that have been recently visited according to their eligibility, when reinforcement is received. 
The backward TD($\lambda$) keep an eligibility trace for every state $s$ and update value $V(s)$ for every state $s$
\begin{center}
$V(s)\leftarrow V(s)+\alpha(R_{t+1}+\gamma V(S_{t+1})-V(S_t))E_t(s),$
\end{center} 
while more details are covered by \cite{Kaelbling1996}, \cite{Dayan1992} and \cite{Dayan1994}.

\subsection{Model-Free Control}
Model-free control stands for optimizing the value function of an unknown MDP. Previously, we discussed how to evaluate a policy through value functions, while in this section, we focus on how to improve a policy function. 
It is worth noting that improving a policy over $V(s)$ requires the model of MDP, i.e., $\pi'(s)=\argmax_{a\in A} R_s^a+P^a_{ss'}V(s'),$ 
where $P^a_{ss'}$ is denoted as the probability of taking an action $a$ while transiting from state $s$ into state $s'$. 
Therefore, the knowledge of such a probability is required, which is not model-free. 
Instead, we can improve the policy over $Q(s,a):\pi'(s)=\argmax_{a\in A} Q(s,a)$ with a simple idea called the $\epsilon$-Greedy Exploration, in which all $m$ actions are tried to choose the greedy action with probability $1-\epsilon$, and choose an action at random with probability $\epsilon.$ Then, we have
\begin{center}\begin{equation*}\pi(a|s)=\begin{cases} \epsilon/m+1-\epsilon \text{ if } a^*=\argmax_{a\in A} Q(s,a)\\
\epsilon/m \text{ otherwise}.\end{cases}\end{equation*}
\end{center}
It has been proved that for any $\epsilon$-greedy policy $\pi$, the $\epsilon$-greedy policy $\pi'$ with respect to $q_\pi$ is an improvement, i.e., $v_{\pi'}(s)\geqslant v_{\pi}(s)$. Thus, we can apply TD to $Q(S,A)$ by using a $\epsilon$-greedy policy improvement and update at every time-step. A naive thought is to use TD(0): $Q(S,A)\leftarrow Q(S,A)+\alpha(R+\gamma Q(S',A')-Q(S,A)),$ where $R$ is the immediate reward obtained from the state $S$ by taking action $A$. This algorithm is also called SARSA On-Policy Control.\\

Similar to the state-value function, we define the n-step $Q$-return as 
$$q_t^{(n)}=R_{t+1}+\gamma R_{t+2}+\dots+\gamma^{n-1}R_{t+n}+\gamma^n Q(S_{t+n}),$$
and use the weight $(1-\lambda)\lambda^{n-1}$ to define the Forward SARSA($\lambda$) \begin{center} $Q(S_t,A_t)\leftarrow Q(S_t,A_t)+\alpha(q_t^{\lambda}-Q(S_t,A_t)).$\end{center}
Following the same logic, we use one eligibility trace for each state-action pair, where $E_0(s,a)=0$ and $E_t(s,a)=\gamma\lambda E_{t-1}(s,a)+\textbf{1}(S_t=s, A_t=a).$  The backward SARSA($\lambda$) updates are then defined as
\begin{center}
$Q(s,a)\leftarrow Q(s,a)+\alpha(R_{t+1}+\gamma Q(S_{t+1},A_{t+1})-Q(S_t,A_t))E_t(s,a).$\end{center} 

\subsection{Q-Learning}
In the so-called off-policy learning, we evaluate some other target policy $\pi(a|s)$ to compute $q_\pi(s,a),$ while following our behavior policy $\mu(a|s),$ and two different policies are used in the policy improvement process. The target policy is used to estimate the value functions, and the behavior policy is used to control the improvement process. Since such a consideration of the off-policy learning is based on the action-values $Q(s,a),$ this algorithm is called Q-learning. Specifically, we allow both behavior and target policies to improve by applying a greedy algorithm to the target policy $\pi$ w.r.t. $Q(s,a),$ and applying $\epsilon$-greedy to the behavior policy w.r.t. $Q(s,a)$. The target policy is a greedy search
\begin{center}
$\pi(S_{t+1})=$arg$\max\limits_{a'}Q(S_{t+1},a').$
\end{center}
Since we have the following sequence of equality 
\begin{center}
$R_{t+1}+\gamma Q(S_{t+1},A')=R_{t+1}+\gamma Q(S_{t+1},\text{arg}\max\limits_{a'}Q(S_{t+1},a'))=R_{t+1}+\max\limits_{a'}\gamma Q(S_{t+1},a'),$\end{center} 
we can update $Q(s,a)$ following the policy $\pi$ as follows
\begin{center}
$Q(S,A)\leftarrow Q(S,A)+\alpha(R+\gamma \max\limits_{a'} Q(S',a')-Q(S,A)).$
\end{center} 

\section{Value Function Approximation}\label{VFA}
\subsection{Parameter and Feature Vector}
So far, we assume that the states are discrete variables, in which every state $s$ has an entry $V(s)$ and every state-action pair have an entry $Q(s,a)$. However, in the financial market, the price of equities is characterized by continuous states, which requires us to generalize from limited states to infinite states.   

Since value function maps state/state-action pair to a value, we can build up a function which estimate value functions everywhere. In this case, we create a parameter vector $\textbf{w}$
$\hat{v}(s,\textbf{w}) \approx v_\pi(s)$ or $\hat{q}(s,a,\textbf{w})\approx q_\pi(s,a),$
and update the parameter $\textbf{w}_k$ at step $k$ using a TD learning, such that the associated value function $\hat{v}_k(s_t,\textbf{w}_k)$ or $\hat{q}_k(s_t,a_t,\textbf{w}_k)$ totally depends on $\textbf{w}_k$ that varies step by step.

Nest, Our objective is to find a parameter vector $\textbf{w}$ minimizing the mean-squared error between approximate value $\hat{v}(s,\textbf{w})$ and true value $v_\pi(s)$ 
\begin{center}$J(\textbf{w})= \mathbb E_\pi[(v_\pi(S)-\hat{v}(S,\textbf{w}))^2],$\end{center} 
where $\hat{v}(S,\textbf{w})$ shows that we can approximate the value of $v_\pi(S)$, given a state $S$ and a parameter vector $\textbf{w}$. The remaining is to find a vector that can represent states
\begin{center}$
\textbf{x}(S)=\begin{pmatrix}
\textbf{x}_1(S)\\
\dots\\
\textbf{x}_n(S)\\
\end{pmatrix}$.\end{center}
We can then represent value function by a linear combination of features
\begin{center}$\hat{v}(S,\textbf{w})=\textbf{x}(S)^T\textbf{w}=\sum\limits_{j=1}^n \textbf{x}_j(S)\textbf{w}_j,$\end{center}
and use a stochastic gradient descent method to minimize the mean-squared error. It can be easily found that 
$\triangledown_\textbf{w} \hat{v}(S,\textbf{w})=\textbf{x}(S)$.
Therefore, to find a local minimum of $J(\textbf{w})$, we adjust the parameter $\textbf{w}$ in the direction
 \begin{center}$\Delta \textbf{w}=-\frac{1}{2}\alpha \triangledown_\textbf{w} J(\textbf{w})=\alpha[(v_\pi(S)-\hat{v}(S,\textbf{w}))\triangledown_\textbf{w} \hat{v}(S,\textbf{w})]=\alpha[(v_\pi(S)-\hat{v}(S,\textbf{w}))\textbf{x}(S)],$\end{center} where $\alpha$ is the step-size.
\subsection{Action-Value Function Approximation}
Following the same analogy, we approximate the action-value function
$\hat{q}(S,A,\textbf{w})\approx q_\pi(S,A)$ by minimizing the mean-squared error
$$J(\textbf{w})=\mathbb E_\pi[(q_\pi(S,A)-\hat{q}(S,A,\textbf{w}))^2],$$ 
and then use the stochastic gradient descent to find its local minimum. Again, a feature vector is defined to represent the state and action pair \begin{center}$\textbf{x}(S,A)=\begin{pmatrix}
\textbf{x}_1(S,A)\\
\dots\\
\textbf{x}_n(S,A)\\
\end{pmatrix},$\end{center}
and the value function is represented by a linear combination of features
\begin{center}$\hat{q}(S,A,\textbf{w})=\textbf{x}(S,A)^T\textbf{w}=\sum\limits_{j=1}^n \textbf{x}_j(S,A)\textbf{w}_j.$\end{center} 
Thus, the stochastic gradient descent update are as follows
\begin{eqnarray*}
\Delta \textbf{w}&=&-\frac{1}{2}\alpha \triangledown_\textbf{w} J(\textbf{w})\\
&=&\alpha[(q_\pi(S,A)-\hat{q}(S,A,\textbf{w}))\triangledown_\textbf{w} \hat{q}(S,A,\textbf{w})]\\
&=&\alpha[(q_\pi(S,A)-\hat{q}(S,A,\textbf{w}))\textbf{x}(S,A)],
\end{eqnarray*} 
where $\alpha$ is the step-size.
We substitute a target for $q_\pi(S,A),$ and update it by 
$E_0(s,a)=0$ and $E_t(s,a)=\gamma\lambda E_{t-1}(s,a)+\textbf{1}(S_t=s, A_t=a).$
In the framework of approximation, the updating rule is 
$E_t=\gamma \lambda E_{t-1}+\triangledown_\textbf{w}\hat{q}(S_t,A_t,\textbf{w})=\gamma \lambda E_{t-1}+\textbf{x}(S,A),$ 
and
\begin{center}
$\Delta \textbf{w}=\alpha(R_{t+1}+\gamma \hat{q}(S_{t+1},A_{t+1},\textbf{w})-\hat{q}(S_t,A_t,\textbf{w}))\textbf{x}(S_t,A_t)E_t.$\end{center}

\subsection{LSTD}
The gradient descent optimization of $\textbf{w}$ works in general, including non-linear value function approximation. However, when the approximation is linear, we can solve the least-squares solution directly, by assuming that at the minimum of $J(\textbf{w})$, the expected update should be zero, i.e., $\mathbb E_\pi[\Delta\textbf{w}]=0$. 
We then obtain 
\begin{align*}&\alpha\sum\limits_{t=1}^T\textbf{x}(s_t)(v_t^\pi-\textbf{x}(s_t)^T\textbf{w})=0,\\
&\sum\limits_{t=1}^T\textbf{x}(s_t)v_t^\pi=\sum\limits_{t=1}^T\textbf{x}(s_t)\textbf{x}(s_t)^T\textbf{w},\\
&\textbf{w}=(\sum\limits_{t=1}^T\textbf{x}(s_t)\textbf{x}(s_t)^T)^{-1}\sum\limits_{t=1}^T\textbf{x}(s_t)v_t^\pi,
\end{align*}
where the true values $v_t^\pi$ are unknown, and our training data is noisy samples of $v_t^\pi$ in practice. 
For LSTD(0), the $\textbf{w}$ is solved by
\begin{center}$\textbf{w}=\bigg(\sum\limits_{t=1}^T\textbf{x}(S_t)(\textbf{x}(S_t)^T-\gamma\textbf{x}(S_{t+1}))^T\bigg)^{-1}\sum\limits_{t=1}^T\textbf{x}(S_t) R_{t+1},$\end{center} 
while for LSTD($\lambda$), the $\textbf{w}$ is solved by
\begin{center}
$\textbf{w}=\bigg(\sum\limits_{t=1}^TE_t(\textbf{x}(S_t)^T-\gamma\textbf{x}(S_{t+1}))^T\bigg)^{-1}\sum\limits_{t=1}^T E_t R_{t+1}.$
\end{center}
In practice, we use LSTDQ algorithm and simply substitute $V(S)$ with $Q(A,S).$ For the LSTDQ(0), the  $\textbf{w}$ is solved by
\begin{center}$\textbf{w}=\bigg(\sum\limits_{t=1}^T\textbf{x}(S_t, A_t)(\textbf{x}(S_t, A_t)-\gamma\textbf{x}(S_{t+1},\pi(S_{t+1})))^T\bigg)^{-1}\sum\limits_{t=1}^T\textbf{x}(S_t, A_t) R_{t+1},$
\end{center} 
and in LSTDQ($\lambda$), we have
\begin{center}$\textbf{w}=\bigg(\sum\limits_{t=1}^T E_t(\textbf{x}(S_t, A_t)-\gamma\textbf{x}(S_{t+1},\pi(S_{t+1})))^T\bigg)^{-1}\sum\limits_{t=1}^T E_t R_{t+1}.$
\end{center} 

\section{Policy Gradient with Actor-Critic}\label{AC}
\subsection{Policy Gradient}
In this section, we use $\textbf{w}$ and $\theta$ to parameterize the value function and the policy function, respectively. 
Our goal is to find the optimal $\theta$, based on a policy $\pi_\theta(s,a)$ with parameter $\theta$. 
In terms of measuring the quality of a policy $\pi_\theta$, we can use the mean value 
\begin{center}
$J_{avV}(\theta)=\sum\limits_s d^{\pi_\theta}(s)V^{\pi_\theta(s)},$
\end{center} 
or the mean reward per time-step
\begin{center}
$J_{avR}(\theta)=\sum\limits_s d^{\pi_\theta}(s)\sum\limits_a\pi_\theta(s,a)R_s^a,$
\end{center} 
where $d^{\pi_\theta}(s)$ is a stationary distribution of Markov Chain for $\pi_\theta$. 
We then need to find $\theta$ that maximizes $J(\theta)$. Similar to the previous section, the policy gradient algorithm search for a local maximum in $J(\theta)$ by ascending the gradient of the policy, w.r.t. parameters $\theta$ 
\begin{center}$\Delta\theta=\alpha \triangledown_\theta J(\theta),$\end{center} 
where $\triangledown_\theta J(\theta)$ is the policy gradient, and $\triangledown_\theta J(\theta)=\begin{pmatrix}
\frac{\partial J(\theta)}{\partial \theta_1}\\
\dots\\
\frac{\partial J(\theta)}{\partial \theta_n}\\
\end{pmatrix}.$
We now compute the policy gradient analytically using the likelihood ratios trick 
\begin{align*}\triangledown_\theta \pi_\theta(s,a)&=\pi_\theta(s,a)\frac{\triangledown_\theta \pi_\theta(s,a)}{\pi_\theta(s,a)}\\
&=\pi_\theta(s,a)\triangledown_\theta \log \pi_\theta(s,a),\end{align*} 
where 
$\triangledown_\theta \log \pi_\theta(s,a)$ is a score function. 
Finally, we apply policy gradient theorem for the two $J(\theta)$ functions and obtain the gradient policy as
\begin{center}$\triangledown_\theta J(\theta)=\mathbb E[\triangledown_\theta \log \pi_\theta(s,a) Q^{\pi_\theta}(s,a)],$\end{center}
while more details can be referred to \cite{Sutton2000}. 

\subsection{Actor-Critic}
We now combine the policy parametrization with the action-value function parametrization, where we use a \emph{critic} to estimate the action-value function $Q_\textbf{w}(s,a)\approx Q^{\pi_\theta}(s,a)$ and update the action-value function parameter $\textbf{w}.$ Next, we update the policy parameter $\theta$ in the direction of our approximated action-value function and follows 
\begin{center}$\triangledown_\theta J(\theta)\approx \mathbb E_{\pi_\theta}[\triangledown_\theta \log \pi_\theta(s,a) Q_\textbf{w}(s,a)],$\end{center}
and 
\begin{center}$\Delta\theta=\alpha \triangledown_\theta \log \pi_\theta(s,a) Q_\textbf{w}(s,a).$\end{center}

\subsection{Advantage Actor-Critic}
One of the disadvantages of the policy gradient method is that it usually has a large variance. To reduce the variance, we subtract a baseline function from the policy gradient to reduce the variance without changing the expectation (\cite{Grondman2012} and \cite{Sutton2000}). A good baseline function can be $V^{\pi_\theta}(s)$, and hence, we have 
\begin{center}$\triangledown_\theta J(\theta)= \mathbb E_{\pi_\theta}[\triangledown_\theta \log \pi_\theta(s,a) (Q^{\pi_\theta}(s,a)-V^{\pi_\theta}(s))],$\end{center}
where $Q^{\pi_\theta}(s,a)-V^{\pi_\theta}(s)$ is called the advantage function. 
In fact, the TD error of $V^{\pi_\theta}(s)$ is an unbiased estimate of the advantage function, and we can use the TD error to compute the policy gradient $\triangledown_\theta J(\theta)= \mathbb E_{\pi_\theta}[\triangledown_\theta \log \pi_\theta(s,a) (r+\gamma V^{\pi_\theta}(s')-V^{\pi_\theta}(s))].$ 
In practice, we can use the approximated TD error, and obtain the policy gradient as
\begin{center} $\triangledown_\theta J(\theta)= \mathbb E_{\pi_\theta}[\triangledown_\theta \log \pi_\theta(s,a) (r+\gamma V_v(s')-V_v(s))],$\end{center} 
and 
\begin{center}$\Delta\theta=\alpha \triangledown_\theta \log \pi_\theta(s,a)(r+\gamma V_v(s')-V_v(s)),$\end{center} 
where $v$ is the set of critic parameters. 
We apply TD(0) as mentioned and obtain
\begin{center}$\Delta\theta=\alpha \triangledown_\theta \log \pi_\theta(s,a)(r+\gamma V_v(s_{t+1})-V_v(s_t)).$\end{center} 
As in the backward-view of TD($\lambda$), we can aplly the eligibility trace and obtain the following updating rule
\begin{center}$E_{t}=\lambda E_{t-1}+\triangledown_\theta \log \pi_\theta(s,a),$\end{center} 
and \begin{center}$\Delta\theta=\alpha E_t(r_{t+1}+\gamma V_v(s_{t+1})-V_v(s_t))$.\end{center}

\section{Conclusion}
In this technical report, we summarized the basic framework of the RL and then went through the development of the algorithms including TD, Q-learning, and Actor-Critic. We also introduced how to use value function approximation to evaluate the state value, policy value, and state-action pair value. With the concrete understanding of the framework of RL, we will be able to apply RL into the field of Quantitative Trading. In recent years, most of the published works relative to this topic are based on dynamic programming, TD learning \cite{Moody1998} \cite{TD} or Q learning \cite{Q}. These methods left the policy gradient method out. On the other hand, some other work such as \cite{Moody2001} only focused on the policy gradient method without using value functions. However, in our paper, we introduced another class of RL algorithm that combines both value function approximation and the policy gradient method, namely, the Actor-Critic algorithm. Therefore, one possible future research direction is to compare the performance of these different types of algorithms on real datasets. Another research direction might be boosting the RL algorithm itself. For example, it would be interesting to try to prove the convexity of the value functions and apply the Interior-point Algorithm while doing policy gradient search. 

In regards to the application of RL in quantitative trading, an agent must also have its own interpretation of the information set, specifically, the external/market variables. Consequently, a natural question to ask is what factors should we take into consideration to fully describe the environment while keeping them simple. In addition, calculating the exact Sharpe Ratio at every time step is extremely expensive and we might need to use the differential Sharpe Ratio for approximation. What would be the risk-adjusted return using RL algorithm taking all of these into account? Answering these research questions is our next goal and we will address them in our future work.

\end{document}